\documentclass{emulateapj}
\shorttitle{THERMAL X-RAY EMISSION FROM LONG GRBS}
\shortauthors{SUZUKI\&SHIGEYAMA}
\begin{document}
\title{EARLY THERMAL X-RAY EMISSION FROM LONG GAMMA-RAY BURSTS AND THEIR CIRCUMSTELLAR ENVIRONMENTS}
\author{AKIHIRO SUZUKI\altaffilmark{1} and TOSHIKAZU SHIGEYAMA\altaffilmark{2}}
\altaffiltext{1}{Center for Computational Astrophysics, National Astronomical Observatory of Japan, Mitaka, Tokyo, 181-8588, Japan.}
\altaffiltext{2}{Research Center for the Early Universe, School of Science, University of Tokyo, Bunkyo-ku, Tokyo, 113-0033, Japan.}
\begin{abstract}
We performed a series of hydrodynamical calculations of an ultra-relativistic jet propagating through a massive star and the circumstellar matter to investigate the interaction between the ejecta and the circumstellar matter. 
We succeed in distinguishing two qualitatively different cases in which the ejecta are shocked and adiabatically cool. 
To examine whether the cocoon expanding at subrelativistic speeds emits any observable signal, we calculate expected photospheric emission from the cocoon. 
It is found that the emission can explain early thermal X-ray emission recently found in some long gamma-ray bursts. 
The result implies that the difference of the circumstellar environment of long gamma-ray bursts can be probed by observing their early thermal X-ray emission. 
\end{abstract}
\keywords{gamma-ray burst: general -- radiation mechanisms: thermal -- shock waves --  supernovae: general}

\section{INTRODUCTION\label{intro}}
Since the discovery of gamma-ray bursts (GRBs), numerous studies have been done to understand their progenitors, the mechanism to produce their highly energetic emission, and the central engine \citep[see, e.g.,][for review]{1999PhR...314..575P,2006RPPh...69.2259M}. 
It is currently known that long GRBs are triggered by the gravitational collapse of massive stars.
The spatial and temporal coincidence of GRB 980425 and SN 1998bw\citep{1998Natur.395..670G} has revealed the connection between long GRBs and a special class of  type Ic supernovae (broad lined type Ic SNe), i.e., the firmly established SN-GRB connection \citep[see, e.g.,][]{2006ARA&A..44..507W}. 
For example, well-known GRBs associated with SNe are GRB030329/SN 2003dh \citep{2003Natur.423..847H,2003ApJ...591L..17S}, GRB 060218/SN 2006aj \citep{2006Natur.442.1008C,2006Natur.442.1011P,2006Natur.442.1018M}, GRB 100316D/SN 2010bh \citep{2011ApJ...740...41C,2012ApJ...753...67B,2012A&A...539A..76O}. 
The increasing number of detected samples of GRB-associated SNe has enabled us to investigate their circumstellar environments. 

Especially, whether the circumstellar matter (CSM) of the progenitor is dilute or dense is of particular interest, because it is expected that the CSM interacts with the ejecta and results in producing high-energy emission. 
The CSM may originate from the stellar material ejected prior to the explosion as a wind or the common envelope if the progenitor of the GRB was in a binary system \citep{2004ApJ...607L..17P}. 

Recently, it is reported that thermal components are found in X-ray spectra of some long GRBs, which are taken by {\it Swift} satellite 100-1000 seconds after the trigger \citep{2006Natur.442.1008C,2011MNRAS.411.2792S,2011MNRAS.416.2078P,2012MNRAS.427.2950S,2012MNRAS.427.2965S}. 
The component is seen as an excess superposed on a power-law non-thermal component that is usually attributed to synchrotron emission from the forward shock, i.e., the afterglow emission. 
Spectral analyses reveal that the component can be fitted by a single blackbody spectrum with  temperature of $k_\mathrm{B}T=0.1$-$0.9$ keV \citep[see,][]{2012MNRAS.427.2950S}. 
The luminosity ranges from $10^{45}$ to $10^{49}$ erg\ s$^{-1}$. 
The contribution of the thermal emission to the total X-ray flux is typically a few \% up to several 10 \%.   
The emitting radii inferred from the fitting results are $10^{12\mathrm{-}13}$ cm, which are much larger than the typical radius of the progenitor star $\la 10^{11}$ cm. 
Their durations are several 100 seconds, up to 1000 seconds for the longest case, GRB 060218, which is classified as a low luminosity GRB associated with a supernova SN 2006aj. 
The number of GRBs whose spectra exhibit the thermal component now reaches several dozens \citep[see,][]{2012MNRAS.427.2950S,2012MNRAS.427.2965S}. 
%%%%%%%%%%%%%%%%%%%%%%%%%
\begin{figure*}[tbp]
\begin{center}
\includegraphics[scale=0.8]{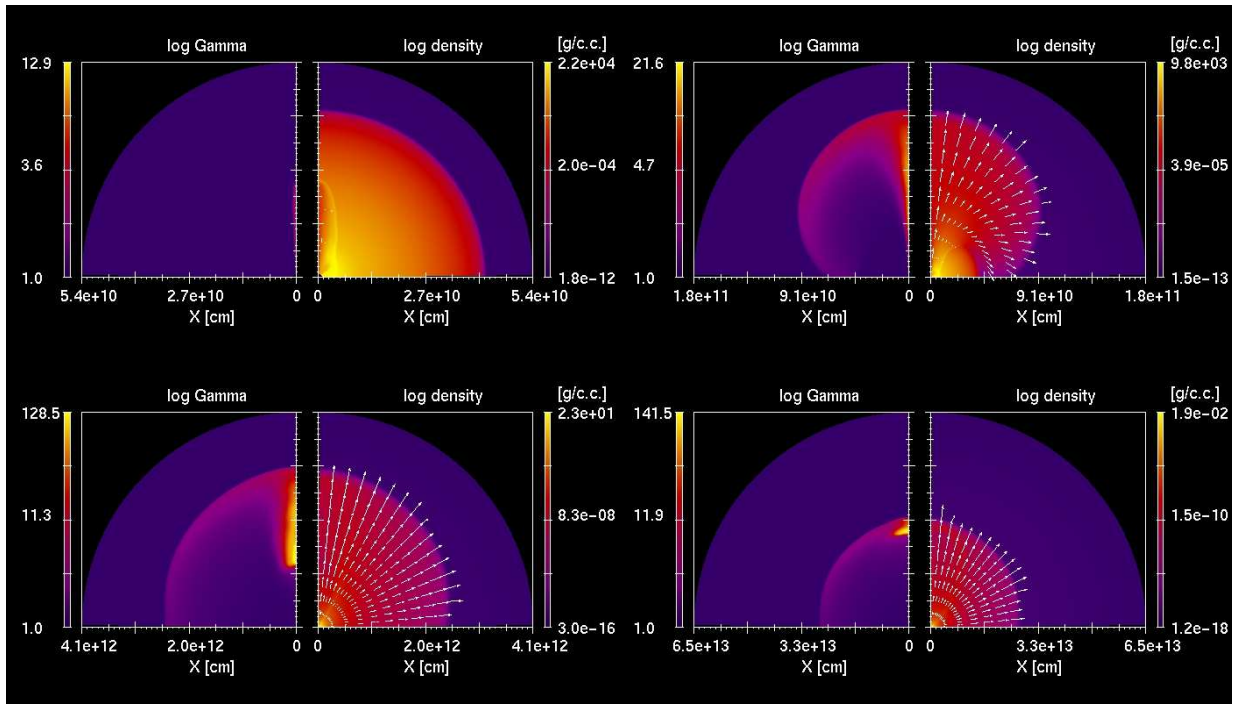}
\caption{Color-coded Lorentz factor and density distributions at $t=6$ (top left), $10$ (top right), $100$ (bottom left), and $10^3$ (bottom right) s for the model with $\dot{M}=10^{-7}\ M_\odot$ yr$^{-1}$.}
\label{figure_snapshots}
\end{center}
\end{figure*}
%%%%%%%%%%%%%%%%%%%%%%%

Several models to explain this emission component have been presented. 
As an example, it is proposed that the supernova shock breakout can be responsible for the emission of some GRBs \citep[e.g.,][]{2007ApJ...667..351W,2007MNRAS.375..240L}. 
On the other hand, for GRB 060218/SN 2006aj, it is pointed out that the radiated energy and the inferred emitting radius are too large to ascribe the emission to the supernova shock breakout from the progenitor surface \citep[e.g.,][]{2007MNRAS.382L..77G}. 
Therefore, some authors ascribe the large emitting radius to the presence of a stellar wind with a high mass-loss rate. 
In this model, the shock emerges from the photosphere located in the wind. 
Another proposed model is the cocoon emission. 
The cocoon is a hot plasma resulting from the interaction between the jet and the stellar material. 
It emerges from the star at the same time the collimated jet penetrates the stellar surface and then expands spherically at mildly relativistic speeds. 
\cite{2006ApJ...652..482P} investigated emission from the cocoon by combining a numerical radiative transfer calculation with an analytical treatment of the dynamical evolution of the cocoon. 
While their model is easy to treat, it is necessary to check whether some parameters used there, such as, the total energy of the cocoon, are realized in actual situations by using hydrodynamical calculations. 
In particular, by performing hydrodynamical calculations, one can estimate the amount of energy deposited into the cocoon out of the total injected energy in a self-consistent way. 
Furthermore, the large emitting radii inferred from spectral analyses indicate that the emission comes from the region where the CSM is expected to be present. 
If so, the ejecta-CSM interaction must give rise to thermal X-ray emission. 
This effect should also be investigated by hydrodynamical calculations. 
In addition, the cocoon emission might be important as a source of seed photons for inverse Compton to produce high-energy photons with energies of $\sim$ 100 MeV, as pointed out by \cite{2009ApJ...707.1404T}.

In this Letter, to investigate the interaction between the ejected matter and the CSM, we perform special relativistic hydrodynamical calculations of the propagation of a relativistic jet emanating from a massive star in the CSM. 
In Section 2, we describe our method to calculate the evolution of the jet and the interaction with the CSM. 
Results of the hydrodynamical simulations and the expected light curves of the emission from the cocoon are presented in Section 3. 
Finally, in Section 4, we discuss implications from the results and conclude this Letter.  

\section{METHOD\label{method}}

In this section, we briefly explain setups of the hydrodynamical calculations performed in this study. 
The detailed code description is found in \cite{thesis}.  
\subsection{Hydrodynamics}
We perform hydrodynamical calculations of the propagation of an ultra-relativistic jet in a massive star and the subsequent interaction with the CSM by using the special relativistic hydrodynamics code in 2D spherical coordinates $(r,\theta)$ developed by one of the authors. 
In this code, we adopt a mapping procedure, in which the width of the radial zones is doubled as the jet head reaches a fraction ($\sim 0.9$) of the maximum of the radial coordinate, in order to calculate the propagation of the jet till $t\sim 1800$ s. 
Thus, the radial resolution becomes coarser as the time elapses.  
At $t=0$, the radial coordinate ranges from $r=10^9$ cm to $r=10^{11}$ cm. 
At the end of the calculations, the maximum of the radial coordinate reaches $r\sim 6\times 10^{13}$ cm. 
The radial zone is divided into $N_r$ uniform cells and the number $N_r=1024$ is fixed. 
The angular coordinate $\theta$ ranges from $\theta=0$ to $\theta=\pi/2$ and is composed of $N_\theta=256$ uniform cells.

\subsection{Simulation setup}
As a presupernova model, we adopt 16TI model in \cite{2006ApJ...637..914W}, which is commonly used in calculations of collapsar jets. 
In this study, we consider several models to clarify the effect of the ejecta-CSM interaction. 
Since the spatial distribution of the CSM is highly uncertain, we adopt the simplest steady wind model whose density profile is given by,
\begin{equation}
\rho_\mathrm{w}(r)=\frac{\dot{M}}{4\pi r^2v_\mathrm{w}}. 
\end{equation}
The density profile is uniquely determined for a given ratio of the mass-loss rate $\dot{M}$ and the wind velocity $v_\mathrm{w}$. 
In this study, the wind velocity $v_\mathrm{w}$ is fixed to be $1000$ km s$^{-1}$. 
We performed calculations with the mass-loss rates of $\dot{M}=10^{-7}$, $10^{-6}$, $10^{-5}$, $10^{-4},$ and $10^{-3}\ M_\odot\ \mathrm{yr}^{-1}$. 
In the following, we especially focus on the two extreme cases, the models with $\dot{M}=10^{-7}$ and $10^{-3}\ M_\odot\ \mathrm{yr}^{-1}$(hereafter they are referred to as the dense and dilute CSM models). 

The jet is injected from the inner boundary $r=10^9$ cm from $t=0$ to $t=60$ s at a constant energy injection rate by using the same method as the previous works \citep[e.g.,][]{2003ApJ...586..356Z,2007ApJ...665..569M,2011ApJ...732...26M}.  
The parameters specifying the jet injection condition are as follows: the total energy $E_\mathrm{tot}=3\times 10^{52}$ erg, the energy injection rate $\dot{E}=5\times 10^{50}$ erg/s, the opening angle $\theta_\mathrm{j}=10^\circ$, the initial Lorentz factor $\Gamma_\mathrm{0}=5$, and the specific internal energy $\epsilon_0/c^2=20$. 

%%%%%%%%%%%%%%%%%%%%%%%%%
\begin{figure}[tbp]
\begin{center}
\includegraphics[scale=0.4]{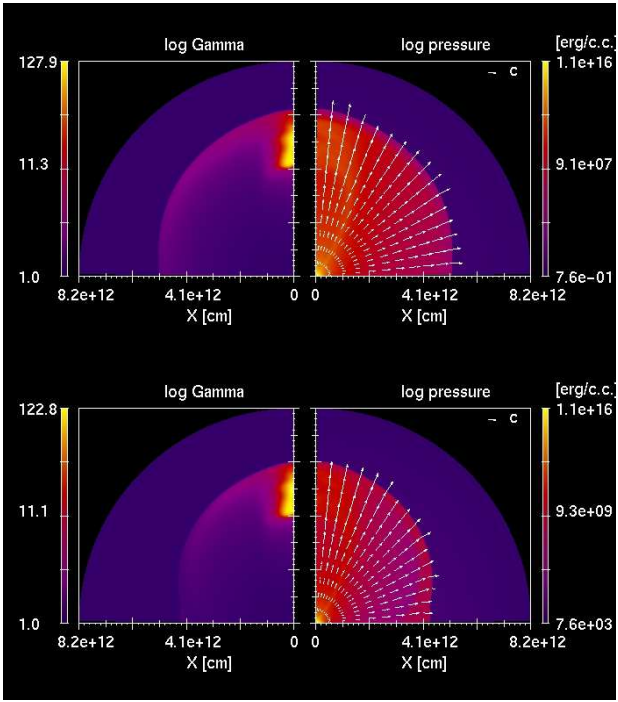}
\caption{Color-coded Lorentz factor and density distributions at $t\sim 200$ s for the model with $\dot{M}=10^{-3}\ M_\odot$ yr$^{-1}$ (lower panel) and  $\dot{M}=10^{-7}\ M_\odot$ yr$^{-1}$  (upper panel).}
\label{figure_compare}
\end{center}
\end{figure}
%%%%%%%%%%%%%%%%%%%%%%%

\section{RESULT\label{result}}
\subsection{Jet dynamics}
A lot of previous works on an ultra-relativistic jet emanating from the progenitor star have been carried out and unveiled the dynamical evolution of the jet, such as, the formation of the recollimation shock and the realization of the well-known fireball solution. 
\citep[e.g.,][]{2003ApJ...586..356Z,2007ApJ...665..569M,2011ApJ...732...26M}. 
Our calculations successfully reproduce and confirm their findings.  
Some snapshots of the spatial distributions of the Lorentz factor and the density of the dilute CSM model are shown in Figure \ref{figure_snapshots}. 
The jet propagates in the interior of the progenitor star and then breaks out, and ejects stellar materials into the circumstellar space. 
As seen in the top right panel, the emergence of a hot material from the jet cavity follows the breakout of the collimated jet. 
The ejecta rapidly expand to form a spherical cocoon as seen in the bottom left panel of Figure \ref{figure_snapshots}. 
It is noteworthy that the cocoon expands at mildly relativistic speeds. 
The appearance and the subsequent expansion of the cocoon have also been reported and investigated by several previous works \citep[see, e.g.,][]{2000ApJ...531L.119A,2002MNRAS.337.1349R,2003ApJ...586..356Z,2005ApJ...629..903L}.

\subsection{Effect of CSM interaction}
\begin{figure}[tbp]
\begin{center}
\includegraphics[scale=0.4]{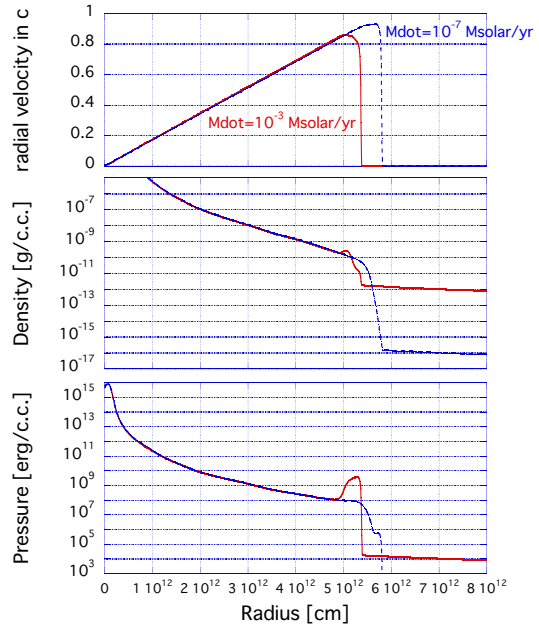}
\caption{Radial profiles along $\theta=45^\circ$ at $t\sim 200$ for the dense CSM model (solid line) and the dilute CSM model (dashed line). 
Each panel represents radial velocity normalized by the speed of light, the density, and the pressure from top to bottom.}
\label{figure_radial}
\end{center}
\end{figure}

Results of the dense and dilute CSM models are compared in Figures \ref{figure_compare} and \ref{figure_radial}. 
Figure \ref{figure_compare} represents the spatial distribution of the Lorentz factor (left) and the pressure (right) at $t\sim 200$ s for the dense CSM model (lower panel) and the dilute CSM model (upper panel).

%The interaction between the ejecta and the CSM is expected to be significant when the energy of the matter swept by the forward shock is comparable to that of the ejecta. 
Near the jet axis ($\theta<10-20^\circ$), no difference between the two models is recognized. 
On the other hand, in the region with large inclination angles ($\theta>20^\circ$), we can see differences between the models. 
Denser CSM reduces the size of the cocoon in comparison with dilute CSM. 
In addition, the pressure distribution shows shell-like structure in the dense CSM. 
 
This difference can also be seen in the radial profiles of some physical variables of the cocoon, as illustrated in Figure \ref{figure_radial}. 
In both cases, the expansion velocities are mildly relativistic as seen in the top panel. 
From the bottom panel showing the pressure profiles, one can see that the reverse shock forms as a result of the cocoon-dense CSM interaction. 
On the other hand, in dilute CSM, the rarefaction wave propagates toward the center in the cocoon. 

%%%%%%%%%%%%%%%%%%%%%%%%%
%%%%%%%%%%%%%%%%%%%%%%%

This is due to the aspherical distribution of the energy deposited into the ejecta. 
Near the jet axis, the energy carried by the jet is too enormous for the CSM to affect the jet propagation. 
On the other hand, the energy deposited into the cocoon component is much smaller than that of the jet. 
The kinetic energy and mass of the cocoon component, which are now defined as those confined in the region outside the star and $\theta>10^\circ$, can be obtained from results of the simulation. 
They are found to be $3\times 10^{50}$ erg and $2\times 10^{-3}$ M$_\odot$ at $t=10$ s and $10^{51}$ erg and $2\times 10^{-2}$ M$_\odot$ at $t=20$ s. 
These values are almost independent of the mass-loss rate, because the dissipation of the kinetic energy of the jet to form the cocoon takes place in the star. 
The kinetic energy and mass of the cocoon increase due to the continuous energy and mass injection by the jet. 
In our calculations, the injection of the jet is terminated at $t=60$ s , which means that the injection of the mass and kinetic energy into the cocoon lasts even after the cocoon begins to expand. 
The kinetic energy of the cocoon is up to a few $\%$ of the total injected energy and thus has a potential for producing thermal X-ray photons with the observed luminosity $\sim 10^{45-48}$ erg s$^{-1}$ for several hundreds seconds.  
On the other hand, the mass of the ultra-relativistic jet component, which is defined as the material with the Lorentz factor larger than 100, is $3\times 10^{-6}$ M$_\odot$, while a substantial fraction ($\sim 10^{51}$ erg) of the injected energy is carried by this component. 

As the radial profiles along $\theta=45^\circ$ at $t=200$ s in Figure \ref{figure_radial} shows, a reverse shock is formed in the dense CSM model.  
The other model with the mass-loss rates $10^{-4}$ $M_\odot\ \mathrm{yr}^{-1}$ also form a reverse shock.  
The energy of the matter ejected immediately after the breakout of the jet from the surface results from the dissipation of a part of the kinetic energy of the jet for the initial several seconds. 
Denoting the dissipated kinetic energy by  $E_\mathrm{dis}$ and the fraction of the internal energy to the total by $\epsilon$, the pressure of the cocoon scales as
\begin{equation}
P_\mathrm{c}\sim \frac{\epsilon E_\mathrm{dis}}{4\pi(v_\mathrm{exp}t)^3},
\end{equation}
where we have assumed that the cocoon is spherically expanding at the velocity $v_\mathrm{exp}$. 
On the other hand, the ram pressure of the CSM behind the forward shock is given by,
\begin{equation}
\rho_\mathrm{w}\Gamma^2c^2\sim \frac{\dot{M}\Gamma_\mathrm{exp}^2c^2}{4\pi v_\mathrm{w}(v_\mathrm{exp}t)^2},
\end{equation}
where $\Gamma_\mathrm{exp}=(1-v_\mathrm{exp}^2/c^2)^{-1/2}$. 
The reverse shock forms when the pressure $P_\mathrm{c}$ of the cocoon becomes comparable to the ram pressure $\rho_\mathrm{w}\Gamma^2c^2$ of the shocked CSM. 
The balance between the pressure of the cocoon and the ram pressure yields the following expression for the time of the reverse shock formation,
\begin{eqnarray}
t&\sim& \frac{\epsilon E_\mathrm{dis}v_\mathrm{w}}
{\dot{M}v_\mathrm{exp}\Gamma_\mathrm{exp}^2c^2}
\sim
10^2
\left(\frac{\epsilon}{5.0\times 10^{-4}}\right)
\left(\frac{E_\mathrm{dis}}{10^{51}\ \mathrm{erg}}\right)\\
&&\hspace{5em}\times
\left(\frac{v_\mathrm{w}}{10^8\ \mathrm{km\ s}^{-1}}\right)
\left(\frac{\dot{M}}{10^{-4}\ M_\odot \mathrm{yr}^{-1}}\right)^{-1}\ \mathrm{s},
\nonumber
%0.0005*1e51/((1-0.8^2)^(-1)*(1e-3*2e33/(365*24*60*60))*3e10^2)*(1e8/(3e10*0.9))=11.68
\end{eqnarray}
where we have derived the final expression by assuming $v_\mathrm{exp}=0.9c$. 
The value of the fraction $\epsilon$ is found from the result of hydrodynamical calculations. 
This rough estimation is consistent with the fact that a reverse shock is observed for models with $\dot{M}=10^{-3}$ and $10^{-4}$ $M_\odot$ yr$^{-1}$ at $t=200$ s and  no reverse shock for models with lower mass-loss rates.   

%%%%%%%%%%%%%%%%%%%%%%%%%
\begin{figure}[tbp]
\begin{center}
\includegraphics[scale=0.45]{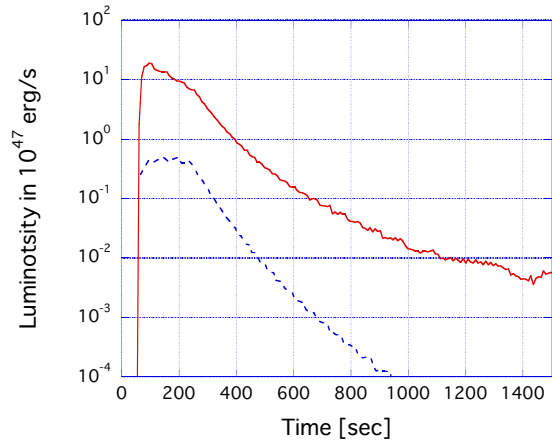}
\caption{Light curves of the photospheric emission calculated for the models with $\dot{M}=10^{-3}\ M_\odot$ yr$^{-1}$ (solid line) and $\dot{M}=10^{-7}\ M_\odot$ yr$^{-1}$ (dashed line).}
\label{figure_emission}
\end{center}
\end{figure}
%%%%%%%%%%%%%%%%%%%%%%%

\subsection{Photospheric emission}
In the following, we investigate whether thermal X-ray emission from GRBs can probe their circumstellar environments. 
We derive the expected light curve and the spectra of thermal X-ray emission from our models by calculating the photospheric emission.

According to \cite{2012MNRAS.427.2950S}, thermal emission with the isotropic luminosity of the order of $10^{47}$ erg s$^{-1}$  and the photon temperature $\sim 0.1-0.9$ keV is observed. 
Illuminated by the radiation, heavy atoms, such as oxygen and carbon, are rapidly photo-ionized. 
The recombination time scale much longer than the ionization time scale keeps those ions fully ionized and the dominant opacity source becomes electron scattering. 
Thus, we calculated the Thomson photosphere from a distant observer along the axis of the jet ($\theta=0$). 
In deriving light curves and spectra, we have assumed that the matter and radiation are strongly coupled and the internal energy density on the photosphere is dominated by that of radiation. 
%It should be noted that the Thomson photosphere is not located on the stellar surface but in the CSM for the case of the dense CSM model. 

%It should be noted that there are some possibilities that the spectrum deviates from the planck function, such as, non-LTE effects and comptonization. 
%Although these effects possibly create photons with the energies much higher than the radiation temperature, they are beyond the scope of this study. 

At first, we briefly consider the properties of the emission. 
Since the ejecta move at mildly relativistic velocities with the Lorentz factor of a few, the relativistic beaming effect strengthen the emission, especially, in the early phase. 
From the top and bottom panels of Figure \ref{figure_radial}, the radiation temperature of the shocked region for the dense CSM can be estimated to be,
\begin{equation}
\Gamma k_\mathrm{B} T_\mathrm{ph}=\Gamma k_\mathrm{B}\left(\frac{3p}{a_\mathrm{r}}\right)^{1/4}\sim 0.1\mathrm{-}0.2\ \mathrm{keV},
\end{equation}
in the observer frame. 
This is consistent with observed values.

\begin{figure}[tbp]
\begin{center}
\includegraphics[scale=0.45]{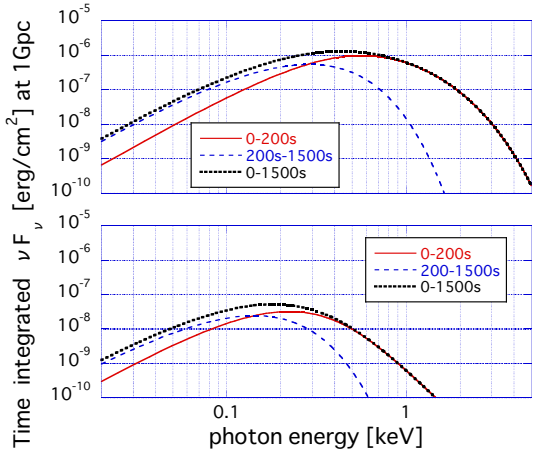}
\caption{Time-integrated $\nu F_\nu$ spectra of the photospheric emission calculated for the models with $\dot{M}=10^{-3}\ M_\odot$ yr$^{-1}$ (upper panel) and $\dot{M}=10^{-7}\ M_\odot$ yr$^{-1}$ (lower panel).
In each panel, $\nu F_\nu$ spectra integrated over $t=0$-$200$ s (solid line), $t=200$-$1500$ s (dashed line), and $t=0$-$1500$ s (dotted line) are plotted.}
\label{figure_spectra}
\end{center}
\end{figure}

The resultant light curves and time-integrated $\nu F_\nu$ spectra of the photospheric emission for both models are presented in Figures \ref{figure_emission} and \ref{figure_spectra}. 
At first, for both models, the photospheric emission is very bright for the first $\sim 200$ sec. 
This is because the cocoon is hot immediately after the emergence from the stellar or wind photosphere. 
After the early phase, the radial velocity at the photosphere decreases as the photosphere moves inward and the cocoon gradually cools. 
This corresponds to the decrease of the luminosity. 
The cocoon cools in different ways for the dense and the dilute CSM. 
As seen in Figure \ref{figure_radial}, the cocoon component is shocked in the dense CSM model. 
The shock converts the kinetic energy of the cocoon into the thermal energy and keeps the shocked region hot. 
As a result, the photospheric emission remains luminous even at $t\sim 1000$ sec. 
In the dilute CSM model, on the other hand, the cocoon adiabatically cools.

Figure \ref{figure_spectra} shows the time-integrated $\nu F_\nu$ spectra for the dense (upper panel) and dilute (lower panel) CSM models.  
In each panel, $\nu F_\nu$ spectra integrated over $t=0$-$200$ s (solid line), $t=200$-$1500$ s (dashed line), and $t=0$-$1500$ s (dotted line) are plotted. 
If we fit a blackbody spectrum with a single temperature to each of these spectra, the temperature is found to be $k_\mathrm{B}T=0.16,$ $0.077$, and $0.13$ keV for the $0$-$200$ s, $200$-$1500$ s, and $0$-$1500$ s spectra of the dense CSM model and $k_\mathrm{B}T=0.061,$ $0.038,$ and $0.051$ keV for the dilute CSM model.
In high-energy part, however, a deviation from the planck function is prominent. 
This shows that each spectrum is actually superposition of blackbody  spectra with different temperatures.

%%%%%%%%%%%%%%%%%%%%%%%%%
%%%%%%%%%%%%%%%%%%%%%%%

\section{DISCUSSION AND CONCLUSIONS}
In this study, we performed hydrodynamical simulations of a jet emerging from a massive star surrounded by the CSM. 
Especially, we focus on the effect of the interaction between the ejecta and the CSM.  
The CSM is assumed to be a steady wind with the wind velocity $v_\mathrm{w}=1000$ km s$^{-1}$ and the mass-loss rates ranging from $\dot{M}=10^{-7}\ M_\odot$ yr$^{-1}$ to $10^{-3}\ M_\odot$ yr$^{-1}$. 
We found that the dynamical behavior of the cocoon, which expands at sub-relativistic speeds, is significantly affected by the ejecta-CSM interaction, while the collimated jet is so energetic that the CSM can not decelerate it even for the dense CSM model. 
In the dense CSM, the cocoon is shocked and thus remains hot even at $\sim$1000 s after the jet injection. 
On the other hand, in the dilute CSM, the cocoon cools adiabatically. 
Furthermore, calculating the photospheric emission from the cocoon, we found that the difference can be detected by observing their early thermal X-ray emission. 

Here we compare our results with the calculation by \cite{2006ApJ...652..482P}. 
They consider emission from a freely expanding spherical cocoon with the initial internal energy of $3\times 10^{51}$, $10^{52}$, and $3\times 10^{52}$ erg. 
On the other hand, the internal energy of the cocoon reproduced in the present calculations is less than these assumed values. 
As a result, for the dilute CSM model, where a freely expanding cocoon is realized, the photon temperature of the photospheric emission is lower than $0.1$ keV. 
For the dense CSM model, a fraction of the kinetic energy of the cocoon can be converted into the internal one by the reverse shock, which leads to a higher photon temperature and brighter photospheric emission. 
This effect is not taken into account by  \cite{2006ApJ...652..482P}. 
%Of course, this value depends on the assumed free parameters characterizing the jet injection condition, such as, the energy injection rate and the jet opening angle. 

From GRB 060218/SN 2006aj, a bright thermal X-ray emission with the temperature $k_\mathrm{B}T\simeq 0.1$-$0.2$ keV is observed even in a few thousand seconds after the trigger. 
\cite{2006Natur.442.1008C} ascribed the long-lived thermal X-ray emission to a high mass-loss rate of $\dot{M}\simeq 3\times 10^{-4}$ $M_\odot$ yr$^{-1}$. 
Our results show that the shocked cocoon is realized at this mass-loss rate. 
Interestingly, some fundamental features of the emission, such as the photon temperature and the long duration, are also reproduced by the photospheric emission from the shocked cocoon . 
Therefore, this event might occur in a dense circumstellar environment. 
However, we cannot reproduce the detailed temporal evolution of the observed photon temperature and the light curve by the present calculations in which the steady wind model is assumed. 
We attribute this discrepancy to the inhomogeneities of the CSM profile as described below. 

We should note that there are some uncertainties in this study. 
In particular, the properties of the photospheric emission calculated in this study are expected to strongly depend on the spatial distribution of the CSM. 
Although we adopted a steady wind model in this study, it may not be realized in actual circumstellar environments of progenitor systems of long GRBs. 
Of course, the slope of the density profile of the wind depends on its mass-loss history prior to the gravitational collapse. 
If the progenitor had been rapidly rotating, the wind could have angular dependence. 
Furthermore, some authors point out that long GRB progenitors have evolved in binary systems \citep[e.g.,][]{2004ApJ...607L..17P}, in which the structure of the material surrounding the system is expected to be much more complicated than single star progenitor cases. 
We regard the influence of the spatial distribution of the CSM on the properties of early thermal X-ray emission as one of future works.

\acknowledgments
We appreciate A. Heger for kindly providing us the presupernova model used in this study.  
This work has been partly supported by  Grant-in-Aid for JSPS Fellows (21$\cdot$1726) of the Ministry of Education, Science, Culture, and Sports in Japan. 
Numerical computations were carried out in part on the Cray XT4 and the middle cluster at the Center for Computational Astrophysics, CfCA, of National Astronomical Observatory of Japan.

\end{document}